\documentclass[12pt]{article}
\usepackage{graphicx}
\topmargin = -\headheight    \voffset = \headheight   \hoffset -5mm
\begin{document}
\addtolength{\voffset}{-1in}
\def\baselinestretch{1.1}   \large
\newcommand{\la}{{\cal L}}

\title{ PHOTON SPLITTING AND PHOTON-PLASMON INTERACTION IN A DENSE MEDIUM}

\vskip 1cm
\author{
V.V. Skalozub\thanks{skalozub@ff.dsu.dp.ua}
~~and A.Y. Tishchenko\thanks{tish@ff.dsu.dp.ua}\\
{\it Dnepropetrovsk State University, Ukraine}}
\date{}

\maketitle

\vskip 1cm
{\small Three-photon vertex in a dense degenerated plasma is calculated.
It is discovered the polarization tensor has the longitudinal part 
which makes possible an interaction between transverse and longitudinal modes 
in the medium. Using dispersion relations it is shown that the  only 
kinematically permitted type of the photon splitting is the decay of the transverse photon to two plasmons}

\vskip 1cm

{\bf 1. Introduction}
\vskip .5cm
The photon splitting is the process with minimal rank of nonlinearity in 
the electromagnetic interactions. This kind of reaction is to be realized 
in various astrophysical objects (white dwarves, neutron stars), the early 
Universe and in the heavy ion collision experiments. In all these cases
Furry's theorem is violated because of the presence of the environments 
such as  external fields or dense media. The photon splitting due to a magnetic field has been rather intensively investigated by many authors 
\cite{Adler}. This process was also considered in the 
rest plasma \cite{Melrose} and the uniformly moving medium \cite{Shabad}.
In the latter article the three-photon vertex has been calculated in one-loop 
approximation for the most general case but the final expressions represented in 
terms of complicate integrals, that makes difficult the detailed analysis of the 
corresponding amplitude. Besides, there were some calculation peculiarities that have not been taken into account and therefore must be pointed out. First it is the use of Feynman's 
parametrization.  As it was proved by Weldon \cite{Weldon}, the employing 
the parametrization at finite temperature and chemical potential could 
lead to errors for some ratios of the momentum components $p_0/p$. The second problem arises because of
the covariant decomposition the authors elaborated. This decomposition allows to 
investigate some general properties of the polarization tensor in moving medium but 
the basis chosen automatically excludes a longitudinal part which tends not to be zero in general case, as it will be shown below.  

Taking all this in mind, we carry out another investigation of the three-photon 
interaction in dense cold medium. In Sect.2  we present straightforward calculation (without Feynman's parametrization) of 
the three-photon vertex in one-loop approximation for arbitrary relations 
between the momenta and the frequencies of the photons.  We do not use the covariant 
decomposition and perform the calculations separately for the temporal, the spatial and the
mixed components of the polarization tensor. For the most important range of 
momenta and frequency for the high density limit ($\mu \gg m, \omega, 
|\vec{k}|$)the final results have been obtained in terms of elementary 
functions. Therefore the analytical properties of the vertex can easily be analysed. One of 
the most interesting features discovered is the longitudinal part of the 
polarization tensor which becomes zero when $\omega=|\vec{k}|$. Obviously, 
the existence of the longitudinal part makes it possible the interaction between the
transversal excitations (photons) in medium and the longitudinal ones 
(plasmons) via the three-photon vertex.

In Sect.3 we consider the photon splitting process for real 
particles. To investigate that the corresponding dispersion relations were obtained. Basing on these relations the kinematic 
analysis of the splitting has been done and it was shown that the only 
type of the permitted process is the splitting of photon to two plasmons. Section 4 contains the conclusions.
In Appendix the expressions for the component of the polarization tensor are presented.

\vskip 1cm

{\bf 3. Three-photon vertex in dense medium} 
\vskip .5cm
The neutron stars and cores of white dwarves are the examples of 
degenerate QED plasma where $T \ll \mu$ and therefore Fermi distribution 
function can be well approximated by step function. For such environment
the three-photon tensor  can be expressed in Euclidean space as 
follows
      \begin{eqnarray} \label{PI}
      &&\Pi_{\mu\nu\lambda}(k,k',k'')\delta(k+k'+k'')=
      \Pi^{(1)}_{\mu\nu\lambda}+\Pi^{(2)}_{\mu\nu\lambda}=
      \nonumber \\
      &&=\delta(k+k'+k'')\frac{e^3}{(2\pi)^4}
      Tr\int d^4 p [\gamma_\mu G(p^*+k)\gamma_\nu G(p^*-k')
      \gamma_\lambda G(p^*)+
      \nonumber \\
      &&+\gamma_\mu G(p^*)\gamma_\lambda G(p^*+k')
      \gamma_\nu G(p^*-k)],
      \end{eqnarray}
where $G(p^*)$ - fermion Green function\\
   \begin{equation}                                         \label{}
   G(p^*)=\frac{-i \hat{p}^* +m}{p^2+m^2};
   \end{equation}
and the notations are introduced
   \begin{equation}                                         \label{}
   p^{*}_{\mu}=\left\{ \begin{array}{ll}
                           p_\mu,  & \mbox{ $\mu=1,2,3$} \\
                           p_4-i\mu,  & \mbox{ $\mu=4$}.
                           \end{array}
                  \right.
   \end{equation}
Corresponding Feynman diagrams are pictured n Fig.1. 

      \begin{figure}
      \centering
      \includegraphics[totalheight=3in,angle=270]{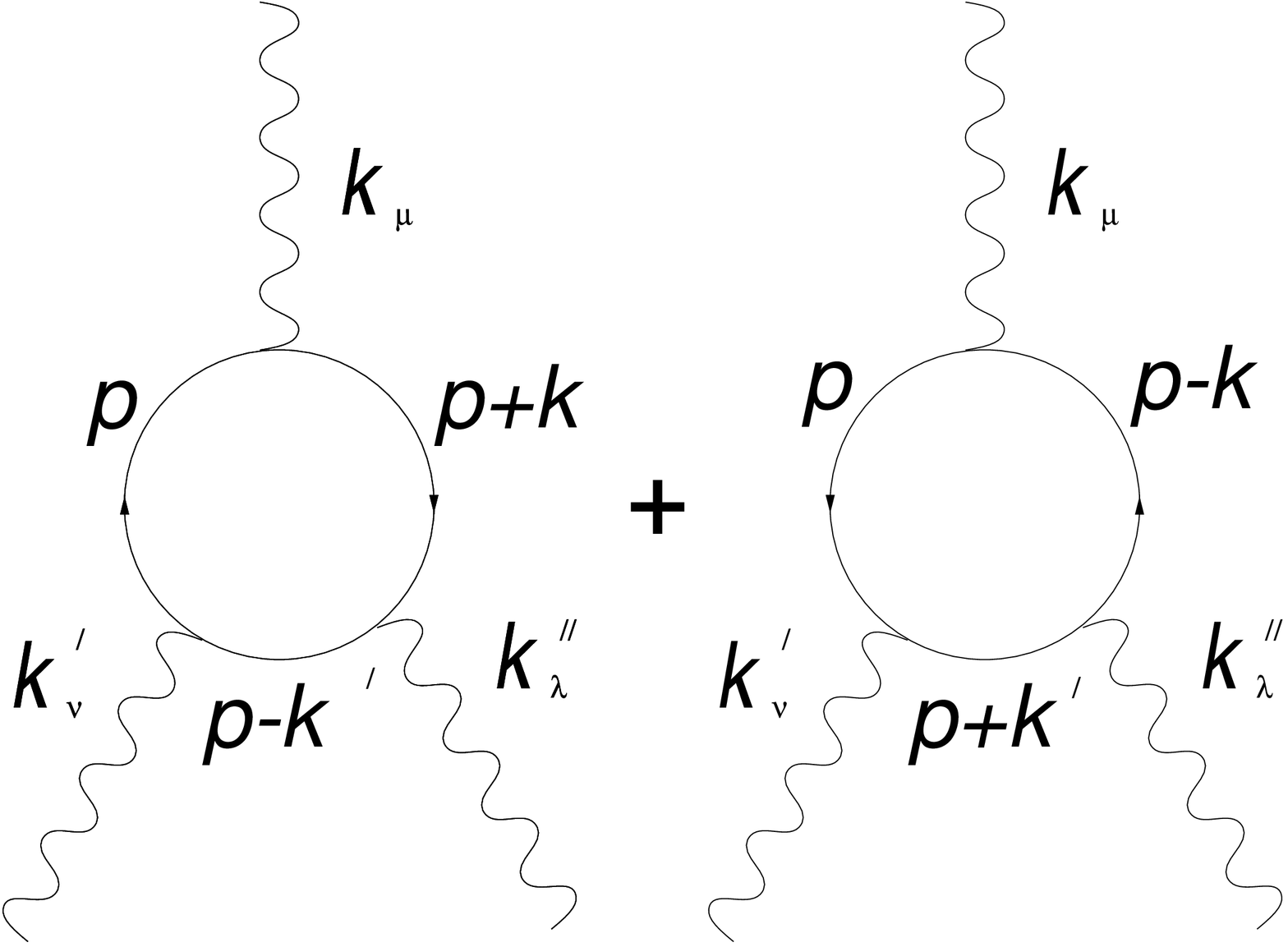}
      \caption{}
      \end{figure}

Because of the reasons mentioned in the introduction  
 we in our calculations will avoid either the Feynman parametrization and the covariant decomposition. Instead, we will perform 
straightforward integration for all the tensor components. 

While integrating over spatial part of the momentum space one faces the
integrals like       
      \begin{equation} \label{int-s}
      \int  \vec{p}~f(p,k,k') ~ d^3p,
      \end{equation}
where $f(p,k,k')$ are certain scalar functions. 

Here we represent $\vec{p}$ as a sum of two components that are transversal 
and parallel to the plane created by $\vec{k},\vec{k'},\vec{k''}$
      \begin{equation} \label{pp}
      \vec{p}=\vec{p_{\perp}}+\vec{p_{\parallel}} 
      \end{equation}

Taking into account the symmetry properties of $f(p,k,k')$ one can easily 
show that only parallel component $\vec{p_{\parallel}}$ contributes to 
integrals (\ref{int-s}). So for further calculations we need basis
consisted only of two independent vectors within reaction plane. 
We chose them as 
      \begin{eqnarray} \label{basis}
      &&\vec{a}=\vec{k}+\vec{k'}, \\      
      &&\vec{b}=-(\vec{k}\times\vec{k'})\times(\vec{k}+\vec{k'})=
      \vec{k}((\vec{k}\vec{k'})+\vec{k'}^2)-
      \vec{k'}((\vec{k}\vec{k'})+\vec{~k~}^2)
      \end{eqnarray}
Expanding $\vec{p_{\parallel}}$ over the normalized basis gives 
      \begin{eqnarray} \label{}
      \vec{p_{\parallel}}=
      \vec{a}\frac{(\vec{p_{\parallel}}\vec{a})}{\vec{~a~}^2} 
      +\vec{b}\frac{(\vec{p_{\parallel}}\vec{b})}{\vec{~b~}^2}=
      \frac{1}{s^2}[(\vec{p_{\parallel}}\vec{k})\vec{Q}(k,k')+
      (\vec{p_{\parallel}}\vec{k'})\vec{Q}(k',k)],
      \end{eqnarray}
where $s\equiv|\vec{k}\times\vec{k'}|$ and
      \begin{eqnarray} \label{q1}
      \vec{Q}(k,k')&=&\vec{k}\vec{k'}^2-\vec{k'}(\vec{k}\vec{k'}),
      \\ \label{q2}     
      \vec{Q}(k',k)&=&\vec{k'}\vec{~k~}^2-\vec{k}(\vec{k}\vec{k'}).
      \end{eqnarray}

The integration over "hard" momenta ($\omega, |\vec{k}| \gg \mu$) makes 
small contribution 
${\displaystyle \sim \mu\Bigg(\frac{\mu}{|\vec{k}|}\Bigg)}$ which can be 
neglected. Integrating over the region ($\omega, |\vec{k}| \ll \mu$) 
(\ref{PI}) we obtain final expressions for high density limit ($\mu \gg m, 
\omega, |\vec{k}|$). These expressions are bulky and placed in 
Appendix.

Having the results for the polarization tensor in analytical form, one can see 
that its components are proportional to chemical potential. It means that 
dense medium may amplify processes corresponding to the three-photon 
interaction. 
 
Next we can note, all the components contain $\theta$-function with
four-momentum squared in its argument. Such a dependence 
restricts
the participation in three-photon interaction in dense medium 
admitting only 
time-like excitations.

One of the most important features of the three-photon vertex can be 
discovered if to calculate the contraction $k_\mu\Pi_{\mu\nu\lambda}$. One can 
convinced that 
      \begin{equation} 
      k_\mu\Pi_{\mu\nu\lambda}\sim k^2.  
      \end{equation}
So, the polarization tensor has a non-zero longitudinal part which becomes 
zero on the mass-shell of free photons. As for physical meaning of this 
result, one can say the existence of the longitudinal part makes possible an 
interaction between transverse (photons) and longitudinal excitations 
(plasmons) via tree-photon vertex in dense medium.

\vskip 1cm

{\bf3. Dispersion relations and photon-plasmon interaction}
\vskip .5cm
All the results in previous part has been made for arbitrary relations 
between momenta and frequencies of external photons. For studying 
the processes with the real particles in medium we need the corresponding 
dispersion relations. This type of calculation has been made by several 
authors for different cases \cite{Sitenko}-\cite{Alther2}.

The relations mostly match our purposes have been obtained  in Ref.
\cite{Alther1} and in our notations look as follows
      \begin{eqnarray} \label{dr1}
      \omega^2-\vec{k}^2=Re \Pi(\omega,\vec{k}),
      \end{eqnarray}
where for transverse excitations we have
      \begin{eqnarray} \label{pt}
      \Pi_T(\omega)=\frac{e^2}{2\pi^2}\frac{k^{3}_{F}}{\mu}
      \Bigg[\Big(\frac{\mu\omega}{k_{F}|\vec{k}|}\Big)^2+
      \frac{1}{2}\frac{\mu\omega}{k_F |\vec{k}|}
      \Bigg(1-\Big(\frac{\mu\omega}{k_F |\vec{k}|}\Big)^2\Bigg)
      \ln\frac{\mu\omega+k_F|\vec{k}|}{\mu\omega-k_F|\vec{k}|}
      \Bigg],
      \end{eqnarray}
and for longitudinal ones
      \begin{eqnarray} \label{pl}
      \Pi_L(\omega, \vec{k})=\frac{e^2}{\pi^2}k_{F}\mu
      \Bigg(1-\frac{\omega^2}{\vec{k}^2}\Bigg)
      \Bigg(1-\frac{1}{2}\frac{\mu\omega}{k_F |\vec{k}|}
      \ln\frac{\mu\omega+k_F|\vec{k}|}{\mu\omega-k_F|\vec{k}|}
      \Bigg).
      \end{eqnarray}
In these formulae $\mu=\sqrt{m^2+k^{2}_{F}}$.

For high density limit ($\mu \gg m, \omega, |\vec{k}|$) the expressions 
(\ref{pt}), (\ref{pl}) give
      \begin{eqnarray} \label{pt}
      \Pi_T(\omega,\vec{k})=\frac{e^2}{2\pi^2}\mu^2
      \Bigg[\Big(\frac{\omega}{|\vec{k}|}\Big)^2+
      \frac{1}{2}\frac{\omega}{|\vec{k}|}
      \Bigg(1-\Big(\frac{\omega}{|\vec{k}|}\Big)^2\Bigg)
      \ln\frac{\omega+|\vec{k}|}{\omega-|\vec{k}|}
      \Bigg], \nonumber \\
      \Pi_L(\omega, \vec{k})=\frac{e^2}{\pi^2}\mu^2
      \Bigg(1-\frac{\omega^2}{\vec{k}^2}\Bigg)
      \Bigg(1-\frac{1}{2}\frac{\omega}{|\vec{k}|}
      \ln\frac{\omega+|\vec{k}|}{\omega-|\vec{k}|}
      \Bigg).
      \end{eqnarray}
Substituting these  in Eq. (\ref{dr1}) we obtain dispersion relations for 
both modes:
      \begin{eqnarray} \label{ft}
      f_T(\omega,\vec{k})&=&\frac{\omega^{2}_{0}}{\vec{k}^2}
      \Bigg(\lambda^2+\frac{\lambda}{2}\Big(1-\lambda^2\Big)
      \ln \Big[\frac{\lambda+1}{\lambda-1}\Big]\Bigg)-\lambda^2+1=0,
      \\                \label{fl}     
      f_L(\omega,\vec{k})&=&\frac{2\omega^{2}_{0}}{\vec{k}^2}
      \Bigg(\frac{\lambda}{2}
      \ln \Big[\frac{\lambda+1}{\lambda-1}\Big]-1\Bigg)-1=0,
      \end{eqnarray}
where $\displaystyle \lambda=\frac{\omega}{|\vec{k}|},~ 
                  \omega_0=\frac{e}{\pi}\mu$.

The dispersion relations obtained are plotted in Fig. 2, where the solid line 
represents the plasmon mode and the dashed line shows the photon mode. As it is seen from the 
 plot, both excitations are time-like and therefore can interact via the
three-photon vertex.

      \begin{figure}
      \centering
      \includegraphics[width=4in]{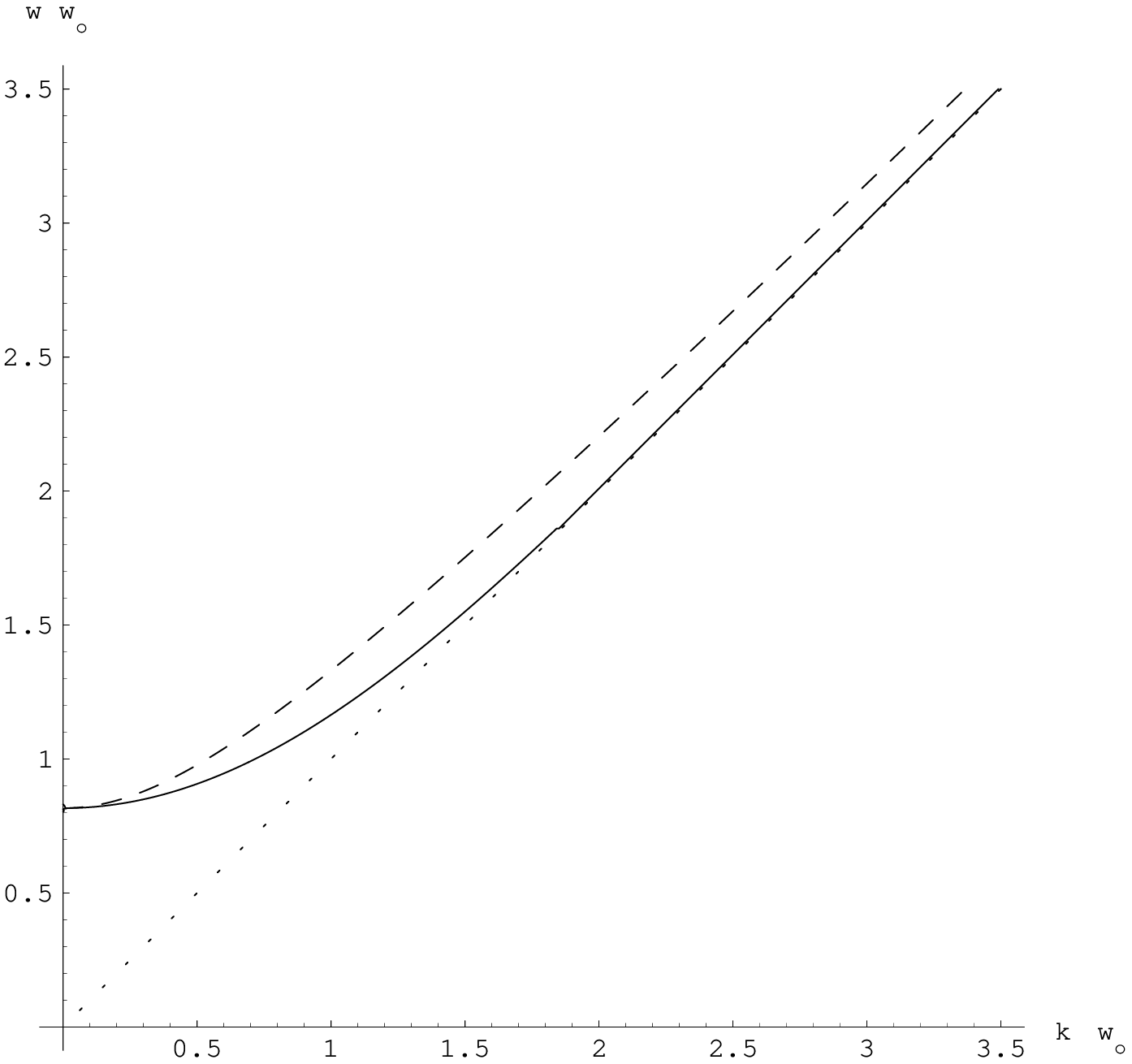}
      \caption{}
      \end{figure}

Having the dispersion relations we are able to analyse the kinematics of splitting of the
real photon in dense media. Let us consider symmetric decay of 
photon with frequency $\omega$ and momentum $\vec{k}$ to two photons with
frequencies $\omega/2$, angle between their momenta $\alpha$ and modulus of 
momenta ${\displaystyle \frac{|\vec{k}|}{2\cos\frac{\alpha}{2}}}$. For this 
case we have
   \begin{equation}                                         \label{}
   \left\{ \begin{array}{l}
           f_T(\omega,|\vec{k}|)=0, \\
           f_T(\frac{\omega}{2},\frac{|\vec{k}|}
            {2\cos\frac{\alpha}{2}})=0.
                           \end{array}
                  \right.
   \end{equation}
Solving this system numerically we find it is has no solutions. Therefore, the 
 photon  splitting to two photons is kinematically forbidden. 

Making the similar consideration for other cases we obtain that the only 
 permitted by kinematics process is the decay of the photon to two plasmons.
The solving of the corresponding equations 
   \begin{equation}                                         \label{}
   \left\{ \begin{array}{l}
           f_T(\omega,|\vec{k}|)=0, \\
           f_L(\frac{\omega}{2},\frac{|\vec{k}|}
            {2\cos\frac{\alpha}{2}})=0
                           \end{array}
                  \right.
   \end{equation}
gives two following results.

First, the photon splitting becomes possible if only its energy exceeds the
minimal value $\approx 2.7 \omega_0$. When the energy of the initial photon 
exactly equals to minimal value the angle between the momenta of plasmons is zero.
 
Besides, if energy of photon becomes greater, the angle increases but its 
value limited approximately by $ 11^{\circ} $.

\vskip 1cm

{\bf 4. Conclusions}
\vskip .5cm
In this paper we have calculated the tree-photon vertex in 
one-loop approximation for dense degenerated medium. The calculations has 
been carried out without Feynman's parametrization and the covariant decomposition.
The results are represented in terms of elementary functions that allows the
detailed analysis of the analytical properties of the polarization tensor.

The most important property discovered is the existence of the longitudinal part 
of the vertex. This part makes possible the interaction between photons and 
plasmons in dense medium. 

Basing on dispersion relations it was shown that all the types of the three-photon 
interaction of real particles are kinematically forbidden but the splitting 
of the photon to two plasmons. This process becomes possible if energy of the
initial photon is larger than a certain minimal value.

\vskip 1cm

{\bf Acknowledgments}

A.T. thanks the Abdus Salam International Center for Theoretical Physics 
where a part of this work has been done for hospitality.


\vskip 1cm
{\bf Appendix}
\vskip .5cm
In this appendix we present the final expressions for the components of 
three-photon polarization tensor

      \begin{eqnarray} \label{Pi444}
      \Pi_{444}&=&-i\frac{e^3\mu}{2\pi}
            \Bigg\{
            \frac{J_1(k,k')}{|\vec{k}||\vec{k'}|}
            (k^2+k'^2+4\omega\omega')+
            \nonumber \\
           &+&
            \frac{J_2(k,k'')}{|\vec{k}||\vec{k''}|}
            (k^2+k''^2+4\omega\omega'')+
            \nonumber \\
           &+&
            \frac{J_3(k',k'')}{|\vec{k'}||\vec{k''}|}
            (k'^2+k''^2+4\omega'\omega'')+
            \nonumber \\
           &+&
            \frac{2\omega I(k)}{|\vec{k}|}+
            \frac{2\omega' I(k')}{|\vec{k'}|}+
            \frac{2\omega'' I(k'')}{|\vec{k''}|}
            \Bigg\},
            \nonumber 
      \end{eqnarray}

      \begin{eqnarray} \label{Pii44}
      \Pi_{i44}=&-&\frac{e^3\mu}{2\pi}    
            \Bigg\{
            \frac{J_1(k,k')}{|\vec{k}||\vec{k'}|}
            \Big[
            2k_i\omega'+
            \nonumber \\      
            &+& 
            \frac{(k'^2+2\omega\omega')}{s^2}
            \big(\omega Q_i(k,k')+\omega' Q_i(k',k)\big)
            \Big]+
            \nonumber \\      
            &+& 
            \frac{J_2(k,k'')}{|\vec{k}||\vec{k''}|}
            \Big[
            2k_i\omega''+
            \nonumber \\      
            &+& 
            \frac{(k''^2+2\omega\omega'')}{s^2}
            \big(\omega Q_i(k,k'')+\omega'' Q_i(k'',k)\big)
            \Big]+
            \nonumber \\      
            &+& 
            \frac{J_3(k',k'')}{|\vec{k'}||\vec{k''}|}
            \Big[
            2k'_i\omega''+2k''_i\omega'
            \Big]+
            \frac{2k_i I(k)}{|\vec{k}|}+
            \nonumber \\      
            &+& 
            \frac{(k'^2+2\omega\omega')}{s^2}
            \Big[
            \frac{I(k')}{|\vec{k'}|}Q_i(k,k')+
            \frac{I(k)}{|\vec{k}|}Q_i(k',k)
            \Big]+
            \nonumber \\      
            &+& 
            \frac{(k''^2+2\omega\omega'')}{s^2}
            \Big[
            \frac{I(k'')}{|\vec{k''}|}Q_i(k,k'')+
            \frac{I(k)}{|\vec{k}|}Q_i(k'',k)
            \Big]
            \Bigg\},
            \nonumber 
      \end{eqnarray}

      \begin{eqnarray} \label{Piij4}
            \Pi_{ij4}=&-&i\frac{e^3\mu}{2\pi}
            \Bigg\{
            \frac{J_3(k',k'')}{|\vec{k''}||\vec{k'}|}
            \Big[
            (2k'_j k''_i-
            \delta_{ij}k'^2)+
            \nonumber \\
           &+&       
            \frac{2\omega''k'_i}{s^2}
            \big(\omega''Q_j(k'',k')+\omega' Q_j(k',k'')\big)
            \Big]+
            \nonumber \\
           &+&       
            \frac{J_2(k,k' )}{|\vec{k''}||\vec{k }|}
            \Big[
            (2k _j k''_i-\delta_{ij}k ^2)+
            \nonumber \\
           &+&       
            \frac{2\omega''k _j}{s^2}
            \big(\omega''Q_i(k'',k )+\omega  Q_i(k ,k'')\big)
            \Big]+
            \nonumber \\
           &+&       
            \frac{2J_1(k,k' )}{|\vec{k'}||\vec{k }|}
            \Big[
            \frac{\omega' k _i}{s^2}
            \big(\omega' Q_j(k',k )+\omega  Q_j(k ,k')\big)+
            \nonumber \\
           &+&       
            \frac{\omega  k'_j}{s^2}
            \big(\omega' Q_i(k',k )+\omega  Q_i(k ,k')\big)
            \Big]+
            \nonumber \\
           &+&       
            \frac{2\omega''k'_i}{s^2}
            \Big[
            \frac{I(k')}{|\vec{k'}|}Q_j(k'',k')+
            \frac{I(k'')}{|\vec{k''}|}Q_j(k',k'')
            \Big]+
            \nonumber \\
           &+&       
            \frac{2\omega''k _j}{s^2}
            \Big[
            \frac{I(k )}{|\vec{k }|}Q_i(k'',k )+
            \frac{I(k'')}{|\vec{k''}|}Q_i(k ,k'')
            \Big]+
            \nonumber \\
           &+&       
            \frac{2\omega' k _i}{s^2}
            \Big[
            \frac{I(k )}{|\vec{k }|}Q_j(k',k )+
            \frac{I(k')}{|\vec{k'}|}Q_j(k ,k')
            \Big]+
            \nonumber \\
           &+&       
            \frac{2\omega  k'_j}{s^2}
            \Big[
            \frac{I(k )}{|\vec{k }|}Q_i(k',k )+
            \frac{I(k')}{|\vec{k'}|}Q_i(k ,k')
            \Big]
            \Bigg\}-
            \nonumber \\
           &-&       
            \frac{2\omega''I(k'')}{|\vec{k''}|}\delta_{ij},
            \nonumber 
      \end{eqnarray}

      \begin{eqnarray} \label{Piijk}
      \Pi_{ijk}&=&i\frac{e^3\mu}{2\pi}
            \Bigg\{
            \frac{J_1(k,k')}{|\vec{k}||\vec{k'}|}
            \Big[
            \frac{(2k_i k'_k-\delta_{ik}k^2)}{s^2}
            \big(\omega Q_j(k,k')+\omega' Q_j(k',k)\big)+
            \nonumber \\      
            &+&    
            \frac{(2k_k k'_j-\delta_{jk}k^2)}{s^2}
            \big(\omega Q_i(k,k')+\omega' Q_i(k',k)\big)
            \Big]+
            \nonumber \\      
            &+&    
            \frac{J_2(k,k'')}{|\vec{k}||\vec{k''}|}
            \Big[
            \frac{(2k_j k''_k-\delta_{jk}k''^2)}{s^2}
            \big(\omega Q_i(k,k'')+\omega'' Q_i(k'',k)\big)+
            \nonumber \\      
            &+&    
            \frac{(2k_i k''_j-\delta_{ij}k'^2)}{s^2}
            \big(\omega Q_k(k,k'')+\omega'' Q_k(k'',k)\big)
            \Big]+
            \nonumber \\      
            &+&    
            \frac{J_3(k',k'')}{|\vec{k'}||\vec{k''}|}
            \Big[
            \frac{(2k'_i k''_k-\delta_{ik}k''^2)}{s^2}
            \big(\omega Q_j(k',k'')+\omega'' Q_j(k'',k')\big)+
            \nonumber \\      
            &+&    
            \frac{(2k''_i k'_j-\delta_{ij}k'^2)}{s^2}
            \big(\omega' Q_k(k',k'')+\omega'' Q_k(k'',k')\big)
            \Big]-
            \nonumber \\      
            &-&
            \frac{2k_i I(k)}{|\vec{k}|}\delta_{jk}-
            \frac{2k'_j I(k')}{|\vec{k'}|}\delta_{ik}-
            \frac{2k''_k I(k'')}{|\vec{k''}|}\delta_{ij}+
            \nonumber \\      
            &+&
            \frac{(2k_i k'_k-\delta_{ik}k^2)}{s^2}
            \Big[
            \frac{I(k')}{|\vec{k'}|}Q_j(k,k')+
            \frac{I(k)}{|\vec{k}|}Q_j(k',k)
            \Big]+
            \nonumber \\      
            &+&
            \frac{(2k_k k'_j-\delta_{jk}k'^2)}{s^2}
            \Big[
            \frac{I(k')}{|\vec{k'}|}Q_i(k,k')+
            \frac{I(k)}{|\vec{k}|}Q_i(k',k)
            \Big]+
            \nonumber \\      
            &+&
            \frac{(2k_i k''_j-\delta_{ij}k^2)}{s^2}
            \Big[
            \frac{I(k'')}{|\vec{k''}|}Q_k(k,k'')+
            \frac{I(k)}{|\vec{k}|}Q_k(k'',k)
            \Big]+
            \nonumber \\      
            &+&
            \frac{(2k_j k''_k-\delta_{jk}k''^2)}{s^2}
            \Big[
            \frac{I(k'')}{|\vec{k''}|}Q_i(k,k'')+
            \frac{I(k)}{|\vec{k}|}Q_i(k'',k)
            \Big]+
            \nonumber \\      
            &+&
            \frac{(2k'_j k''_i-\delta_{ij}k'^2)}{s^2}
            \Big[
            \frac{I(k'')}{|\vec{k''}|}Q_k(k',k'')+
            \frac{I(k')}{|\vec{k'}|}Q_k(k'',k')
            \Big]+
            \nonumber \\      
            &+&
            \frac{(2k'_i k''_k-\delta_{ik}k''^2)}{s^2}
            \Big[
            \frac{I(k'')}{|\vec{k''}|}Q_j(k',k'')+
            \frac{I(k)}{|\vec{k}|}Q_j(k'',k')
            \Big]
            \Bigg\},
            \nonumber 
        \end{eqnarray}
where $s\equiv|\vec{k}\times\vec{k'}|, k^2=\omega^2-\vec{~k~}^2$.

The functions $I,J_1,J_2,J_3$ defined as follows
       \begin{equation}\label{}
      I(k)=\theta(k^2)\ln \Big[\frac{\lambda+1}{\lambda-1}\Big],
            \nonumber 
      \end{equation}

      \begin{eqnarray} \label{J1}
     &&J_1(k,k')=\frac{\theta(k^2)\theta(k'^2)}{B(k,k')}
            \Bigg\{
            2\ln \Big[\frac{\lambda''+1}{\lambda''-1}\Big]+
            \nonumber \\
           &&\hspace{-1.2cm}
            +\ln \Bigg[
            \frac{
            \lambda-\cos\beta-\frac{|{\bf k''}|}{|{\bf k'}|}
            \frac{1}{B(k,k')}
            \Big((1-\lambda \cos\beta)(1+\lambda'')-(\lambda-\cos\beta)^2\Big)
            }
            {\lambda+\cos\beta-\frac{|{\bf k''}|}{|{\bf k'}|}
            \frac{1}{B(k,k')}
            \Big((1+\lambda \cos\beta)(1-\lambda'')-(\lambda+\cos\beta)^2\Big)
            }
            \Bigg]+
            \nonumber \\
           &&\hspace{-1.2cm}
            +\ln \Bigg[
            \frac{
            \lambda'-\cos\alpha-\frac{|{\bf k''}|}{|{\bf k}|}
            \frac{1}{B(k,k')}
            \Big((1-\lambda'\cos\alpha)(1+\lambda'')-(\lambda'-\cos\alpha)^2\Big)}
            {\lambda'+\cos\alpha-\frac{|{\bf k''}|}{|{\bf k}|}
            \frac{1}{B(k,k')}
            \Big((1+\lambda' \cos\alpha)(1-\lambda'')-(\lambda'+\cos\alpha)^2\Big)
            }
            \Bigg]
            \Bigg\},
            \nonumber 
      \end{eqnarray}
where
$\displaystyle{ B(k,k')=\sqrt{\lambda^2+\lambda'^2-
2\lambda\lambda'\cos\gamma-\sin^2\gamma}},\\
\displaystyle{ \lambda=\frac{\omega}{|\vec{k}|},
\lambda'=\frac{\omega'}{|\vec{k'}|},
\lambda''=\frac{\omega''}{|\vec{k''}|}},\\
\alpha=\widehat{\vec{k'}\vec{k''}},
\beta=\widehat{\vec{k}\vec{k''}},
\gamma=\widehat{\vec{k}\vec{k'}}$.

The functions $J_2$ and $J_3$ can be derived from $J_1$ by following
correspondent substitutions in its argument $(k' \leftrightarrow k'', 
\gamma\leftrightarrow\beta), (k \leftrightarrow k'', 
\gamma\leftrightarrow\alpha)$.

\end{document}